\documentclass{Interspeech}

\interspeechcameraready

\providecommand{\paragraph}[1]{\vspace{1em}\noindent\textbf{#1} }

\usepackage{comment}
\usepackage{tabularx}
\usepackage{pifont}
\usepackage{subfig}
\usepackage{adjustbox}
\usepackage{multirow}
\usepackage{multicol}

\usepackage[dvipsnames, table, xcdraw]{xcolor} 
\usepackage{pifont} 
\definecolor{bsRed}{rgb}{0.95, 0.0, 0.0}
\usepackage{hyperref}
\definecolor{bsRed}{rgb}{0.95, 0.0, 0.0}
\definecolor{mygray}{RGB}{128,128,128} 
\definecolor{lightgray}{RGB}{200,200,200}
\usepackage{hyperref}
\usepackage{cite}

\begin{document}


\title{Joint Fullband-Subband Modeling for High-Resolution SingFake Detection
\thanks{*Equal Contribution. }
}

\author[affiliation={1*}]{Xuanjun}{Chen}
\author[affiliation={2*}]{Chia-Yu}{Hu}
\author[affiliation={4}]{Sung-Feng}{Huang}
\author[affiliation={5}]{Haibin}{Wu}
\authorbreak
\author[affiliation={1,3}]{Hung-yi}{Lee}
\author[affiliation={2}]{Jyh-Shing Roger}{Jang}

\affiliation{}{Graduate Institute of Communication Engineering}{National Taiwan University}
\affiliation{}{Department of Computer Science and Information Engineering}{National Taiwan University}
\affiliation{}{NTU Artificial Intelligence Center of Research Excellence}{National Taiwan University}
\affiliation{}{NVIDIA Taiwan}{$^5$Independent Researcher}


\maketitle

\begin{abstract}
Rapid advances in singing voice synthesis have increased unauthorized imitation risks, creating an urgent need for better Singing Voice Deepfake (SingFake) Detection, also known as SVDD. Unlike speech, singing contains complex pitch, wide dynamic range, and timbral variations. Conventional 16 kHz-sampled detectors prove inadequate, as they discard vital high-frequency information. This study presents the first systematic analysis of high-resolution (44.1 kHz sampling rate) audio for SVDD. We propose a joint fullband-subband modeling framework: the fullband captures global context, while subband-specific experts isolate fine-grained synthesis artifacts unevenly distributed across the spectrum. Experiments on the WildSVDD dataset demonstrate that high-frequency subbands provide essential complementary cues. Our framework significantly outperforms 16 kHz-sampled models, proving that high-resolution audio and strategic subband integration are critical for robust in-the-wild detection. 

\keywords{singing voice deepfake detection, subband modeling, high-resolution audio}
\end{abstract}

\section{Introduction}
\label{sec:intro}
Due to the rapid advancement of singing voice synthesis methods, tools such as VISinger \cite{zhang2022visinger} and DiffSinger \cite{liu2021diffsinger} can now generate highly realistic vocals, significantly increasing the risk of unauthorized imitation. This evolution has created an urgent need for robust Singing Voice Deepfake (SingFake) Detection, also known as SVDD, a demand that has led to the establishment of benchmarks like SingFake \cite{zang2023singfake} and the SVDD Challenge \cite{zhang2024svdd_slt}.  
Given the great success of speech deepfake detection, most SVDD research \cite{Nes2Net, Gohari2025audio, phukan2024music, Nguyen-Duc2025comparative, wu2025gasgm, Guragain2024speech} has heavily relied on established methodologies from the speech domain. However, professional singing exhibits acoustic properties that differ fundamentally from standard speech. In particular, singing incorporates intricate harmonic structures and breath-related nuances that extend far into the ultra-high frequency range \cite{high_frequency_energy}. 
Because existing SVDD systems are largely adapted from speech-centric models, they tend to prioritize phonetic cues in the lower spectral regions. Consequently, these systems often overlook the fine-grained spectral ``fingerprints'' in the higher frequencies that are essential for identifying sophisticated singing forgeries.

\begin{figure}[t]
    \centering
    \includegraphics[width=8.5cm]{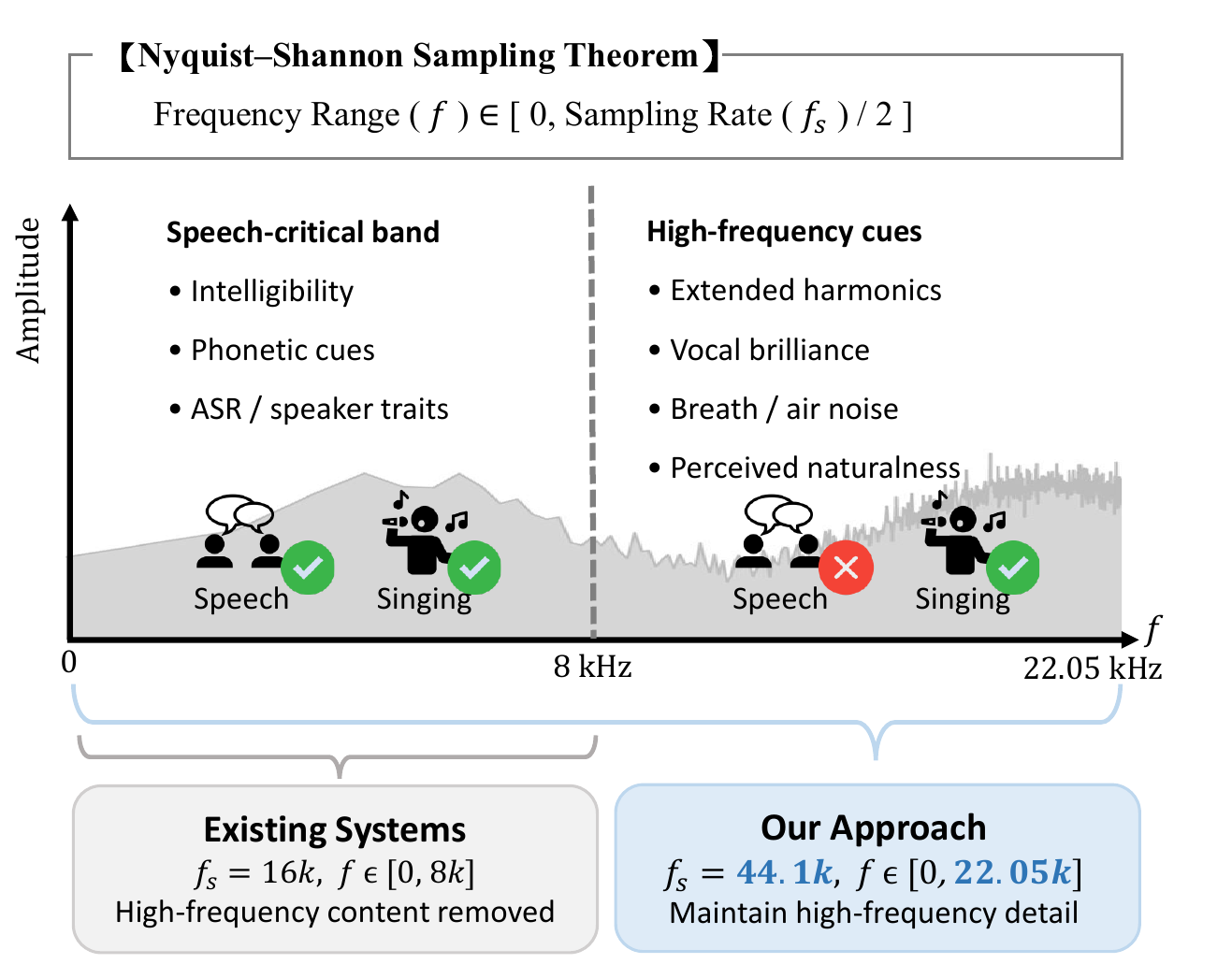}
    \caption{Comparison of audio spectral coverage under different sampling rates. 
    Existing systems typically process 16 kHz sampled audio, restricting them to the speech-critical band (0–8 kHz) and discarding high-frequency details. In contrast, our approach utilizes 44.1 kHz audio to cover the full spectral range (0–22.05 kHz). This preserves extended harmonics and breath textures essential for detecting sophisticated singing forgeries that are mathematically invisible at lower sampling rates. }
    \label{fig:high-low_frequency_comparison}
\end{figure}

Conventional deepfake speech detection typically operates on audio sampled at 16 kHz, a standard primarily optimized for speech intelligibility rather than fullband fidelity. This restricted input resolution creates a significant informational gap, which is largely rooted in the physical constraints defined by the Nyquist–Shannon sampling theorem \cite{Shannon1949}. 
The theorem dictates that the highest representable frequency ($f_{max}$) must satisfy $f_{max} \le f_s/2$, establishing the sampling rate ($f_s$) as the limiting factor for audio resolution, independent of the actual spectral content ($f$) present in the signal. 
According to this theorem, models operating at a restricted sampling rate of $f_s = 16$ kHz encounter an inherent ``spectral ceiling'' at the Nyquist frequency of 8 kHz. 
As illustrated in Figure~\ref{fig:high-low_frequency_comparison}, while this bandwidth is sufficient to capture the speech-critical frequency bands, it inevitably discards the high-frequency cues that are essential for characterizing singing voices. 
In contrast, audio sampled at $f_s = 44.1$ kHz provides a Nyquist frequency of 22.05 kHz. This extended bandwidth encompasses the full audible spectrum and the intricate high-frequency harmonic structures specific to professional singing. 
By maintaining this resolution, high-fidelity recordings preserve fine-grained spectral ``fingerprints,'' such as extended harmonics and breath textures.

The distribution of synthetic artifacts often varies across the frequency spectrum. While prior works in speech anti-spoofing have used subband analysis to isolate these localized anomalies \cite{8917601, chettri2020subbandmodelingspoofingdetection, Xue_2022, 10.1016/j.specom.2023.102988}, their scope remains largely restricted to narrow-band speech where the spectral range is limited. 
The transition to high-resolution SVDD ($0–22.05$ kHz) significantly expands the spectral landscape. The singing voice audio with 44.1 kHz sampling rate exhibits complex pitch variations and rich textures, such as extended harmonics and vocal brilliance, which intuitively may offer critical cues for deepfake detection. However, the precise subband ranges within these high-frequency regions that contribute most to distinguishing synthetic artifacts remain unidentified. Consequently, understanding the individual contributions of these distinct frequency bands in a high-resolution setting is a critical yet largely unexplored problem.

In this paper, we specifically explore how high-resolution audio can be utilized to capture the intricate cues unique to singing voice deepfake detection. Our contributions include:
\begin{itemize}
    \item Through preliminary subband analysis, we reveal a critical limitation: a simple fusion of subband models consistently fails to outperform a dedicated fullband expert. Furthermore, we demonstrate that synthetic artifacts in high-resolution singing voice deepfakes are non-uniformly distributed across the spectrum. 
    These findings led us to emphasize a holistic fullband approach for global cues, while incorporating an enhanced subband focus to capture localized details. 

    \item Building on these insights, we introduce Sing-HiResNet, a joint fullband-subband modeling framework that concurrently capture fullband global context and localized subband artifacts. By implementing and evaluating four specialized fusion strategies, our method systematically quantifies how different spectral perspectives contribute to detection performance. This framework ensures that both speech-critical bands and high-frequency cues are preserved and utilized.

    \item To the best of our knowledge, this work represents the first systematic investigation into high-resolution ($f_s = 44.1$ kHz) SingFake detection, with our Sing-HiResNet achieving state-of-the-art (SOTA) results on the WildSVDD dataset.
\end{itemize}

\section{Related Work}
\subsection{Singing Voice Deepfake Detection}
Singing Voice Deepfake Detection (SVDD) has gained increasing attention as a specialized extension of speech anti-spoofing. Building upon the initial SingFake dataset \cite{zang2023singfake}, the SVDD Challenge 2024 \cite{zhang2024svddchallenge2024singing} expanded the task by introducing two distinct tracks: a controlled setting (CtrSVDD~\cite{zang24_interspeech}) and a in-the-wild setting (WildSVDD). 
The majority of high-performing systems in the challenge leveraged large-scale self-supervised learning (SSL) backbones. For instance, IMS-SCU~\cite{qiu2024wildsvdd} utilized an ensemble including XLS-R and WavLM \cite{chen2022wavlm}, while SingGraph \cite{chen2024singingvoicegraphmodeling} combined MERT \cite{li2024mert} musical representations with wav2vec 2.0 \cite{baevski2020wav2vec} features through graph modeling. While these SSL-based approaches benefit from massive pre-training, they are almost limited to a sampling rate $f_s \le 16$ kHz. 
This constraint inherently discards high-frequency spectral content, which may contain the  ``fingerprints'' necessary to distinguish high-fidelity synthetic singing from human performance.

A notable exception was UNIBS~\cite{10888452}, which employed a lightweight ResNet18~\cite{resnet18} on 44.1 kHz audio. This approach showed that high-resolution spectral information offers a greater performance boost than using SSL-based backbones on downsampled low-resolution audio. 
However, as a purely fullband model, UNIBS overlooks potential frequency-localized discriminative cues. Our work bridges this gap by transitioning from fullband-only modeling to a joint fullband-subband framework, designed to capture both global context and the fine-grained localized artifacts of singing voice deepfakes.

\subsection{Subband Modeling and Integration Strategies}
Subband modeling overcomes the inherent limitations of fullband modeling by isolating localized spectral anomalies that would otherwise be diluted. This approach is effective because synthesis artifacts are often unevenly distributed across the spectrum \cite{8917601, chettri2020subbandmodelingspoofingdetection}. Specifically, while low-frequency harmonics and high-frequency noise residuals provide reliable cues \cite{8917601}, these subtle patterns are easily hidden when the entire spectrum is processed as a single input. By focusing on specific subbands, models can better capture localized anomalies that would otherwise be ignored in a fullband representation. This strategy has therefore been successfully used to capture pitch-related artifacts \cite{Xue_2022} and to improve phase-aware spoof detection \cite{10.1016/j.specom.2023.102988}. 
Nevertheless, prior research has largely focused on narrowband audio ($f_s \le 16$ kHz), leaving the impact of subband modeling on high-resolution signals unexplored.

While isolating these subband cues is essential, the overall performance depends equally on how effectively they are combined with global context. This necessity has led to the development of various subband integration strategies for aggregating localized information. In anti-spoofing, research has moved beyond simple score fusion to dynamic cross-attention mechanisms that weigh subband details against global features \cite{cross_att_2022}. 
However, these integration methods remain largely limited to narrowband speech and fixed spectral partitions.

Our research represents the first systematic exploration of high-resolution SVDD through a multi-scale subband modeling and integration framework. 
Unlike prior works that rely on a single integration mechanism or fixed partitions, we evaluate a hierarchical framework that bridges disparate spectral granularities. 
By comparing strategies ranging from Decision-Level Aggregation to Cross-Expert Distillation, we provide the first systematic investigation into how multi-scale subband experts can be most effectively synergized to detect sophisticated singing voice forgeries.

\section{Proposed Sing-HiResNet Framework}
\label{sec:Sing-HiResNet}

\begin{figure*}[t]
\centering
\includegraphics[width=17cm]{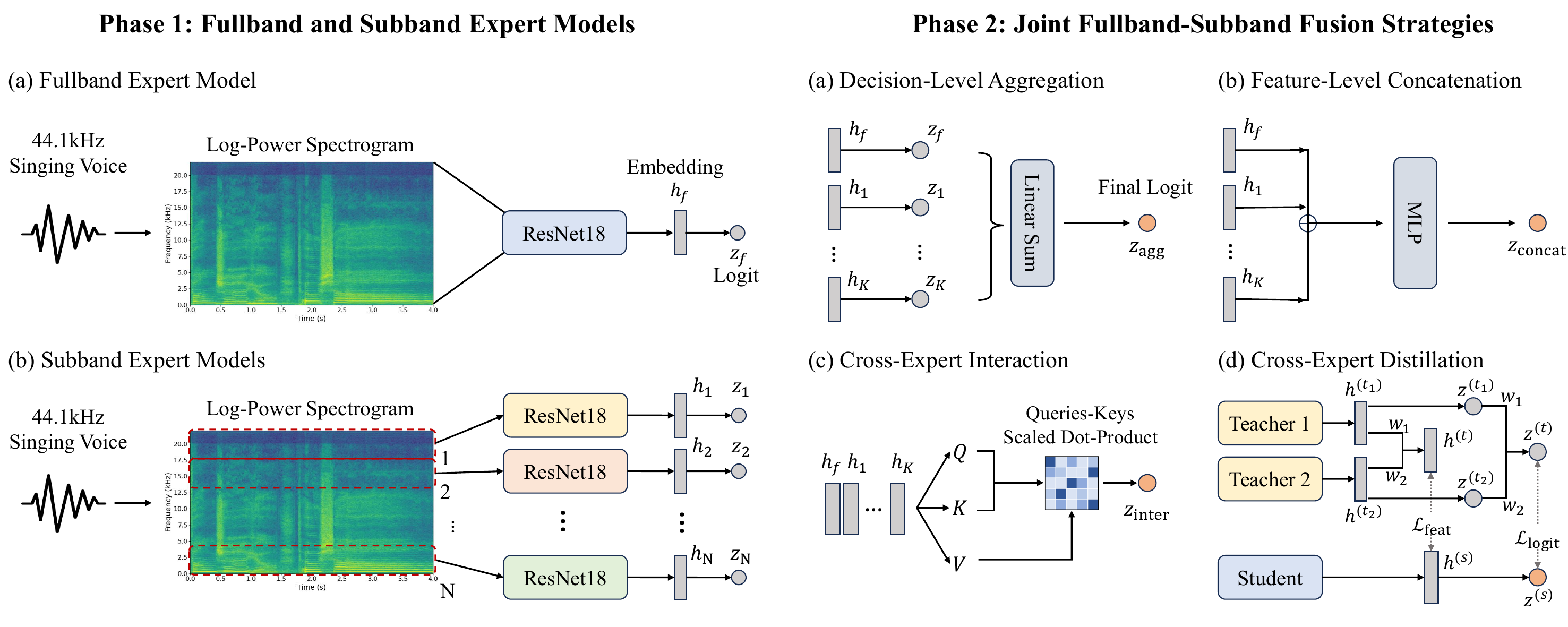} 
\caption{The overview of our proposed Sing-HiResNet framework. The framework is implemented in two stages: Phase 1 establishes the backbone for fullband and subband expert models, while Phase 2 facilitates their integration through various joint fusion processes. }
\label{fig:overview}
\end{figure*}

To better capture the synthesis artifacts in high-resolution singing audio, we propose Sing-HiResNet, a joint fullband-subband framework. This architecture is designed to simultaneously model global spectral patterns and local frequency-specific features. 
As shown in Figure~\ref{fig:overview}, our approach is based on the principle that singing voice deepfake detection requires both a broad view of the fullband spectrum and a detailed check of artifacts within specific subbands. To achieve this, we structure Sing-HiResNet into two interconnected phases:
\begin{itemize}
    \item \textit{Phase 1: Fullband and Subband Expert Models. } This phase have a fullband model that captures global characteristics and subband models that focus on specific frequency ranges.

    \item \textit{Phase 2: Joint Fullband-Subband Fusion Strategies. } Building on the above expert models, this phase integrates the global context with localized artifacts through specific fusion strategies to achieve a more robust detection decision.

\end{itemize}
The following subsections detail our subband partitioning and the specific architecture used for the fusion of different experts.

\subsection{Phase 1: Fullband and Subband Expert Models}
Phase 1 focuses on building feature extraction backbones for both global and local domains. We argue that a single-scale model cannot provide the multi-scale perspective needed for high-resolution audio; therefore, our framework is designed to create a synergy between global context and subband-specific details. 
The fullband model captures broad, long-range spectral dependencies across the entire frequency range. In parallel, a series of subband experts are dedicated to isolating and amplifying subtle, frequency-localized artifacts that are typically obscured in a unified global representation.  
As illustrated in Figure~\ref{fig:overview} (Phase 1), our framework employs two types of expert architectures: a fullband model and multiple subband models. 
Building on the motivation discussed in Section~\ref{sec:intro}, we utilize 44.1~kHz audio to leverage its full 22.05~kHz Nyquist bandwidth ($B_{nyq}$). The fullband model directly processes this complete spectrum to capture global context correlations. Simultaneously, subband expert models are utilized for localized analysis. Following the subband partition method described in previous work \cite{chettri2020subbandmodelingspoofingdetection}, we divide $B_{nyq}$ into $N \in \{1, 2, 4, 8\}$ uniform, non-overlapping segments, assigning each expert model to an equal $1/N$ spectral band as model input.
For instance, by employing $N=2$, we bifurcate the audio to separately analyze vocal components and high-frequency textures, which ensures that fine-grained subband artifacts are precisely captured. 

Each expert model utilizes a ResNet18 model \cite{resnet18, zhang2024svdd_slt} as backbone trained independently to ensure specialization in local artifact detection. 
The original classification layer of ResNet18 is replaced with a projection block producing a 32-dimensional embedding $h$. A lightweight projection head is then appended to map $h$ to the final logit $z$.
By optimizing these backbones individually, we ensure that each expert specializes in its respective spectral range, providing the foundational embeddings and scores required for the subsequent fusion phase. 

\subsection{Phase 2: Joint Fullband-Subband Fusion Strategies} 
The goal of this stage is to fuse the knowledge from both backbones to improve SingFake detection. The fullband model captures the global context and overall structure of the signal, while subband expert models focus on fine-grained artifacts in specific frequency ranges. 
Since subtle forgery cues in high frequencies are often buried by dominant low-frequency components, this fusion ensures that local details are not lost and are instead validated against the global background for a more reliable final decision. 
To facilitate the various fusion strategies detailed below, we first define an selected model pool $\mathcal{P}$ consisting of the fullband model and a subset of subband expert models. 
From this pool, we derive a common set of foundational inputs for any given audio: a 32-dimensional embedding $h_f$ and output logit $z_f$ from the fullband model representing global context, alongside a set of embeddings $\{h_1, \dots, h_K\}$ and logits $\{z_1, \dots, z_K\}$ from the $K$ selected expert models in $\mathcal{P}$ capturing local details. 
Leveraging these multi-scale features, we systematically investigate four joint fullband-subband fusion strategies: (i) decision-level aggregation, (ii) feature-level concatenation, (iii) cross-expert interaction, and (iv) cross-expert distillation. 

\subsubsection{Decision-Level Aggregation}
Starting with a straightforward integration, we implement a parameter-free late fusion strategy that utilizes the logits $z_i$ derived from the selected model pool $\mathcal{P}$. 
This approach aggregates the fullband and selected subband expert models into a single predictive logit:
\begin{equation}
z_{\text{agg}} = \frac{1}{|\mathcal{P}|} \sum_{i \in \mathcal{P}} z_i,
\end{equation}
where $z_i$ denotes the detection logit from the $i$-th model in the pool. 
By treating each model as an equal voter, this strategy captures the inherent complementarity between global context and localized spectral artifacts within a unified prediction framework, requiring no additional training.  
However, as fullband or subband contributions vary across audio samples, unweighted aggregation may introduce interference from less informative spectral regions. 
This motivates the learnable integration strategies described below to optimize cross-expert weights. 

\subsubsection{Feature-Level Concatenation}
To move beyond unweighted logit fusion, we implement a feature-level strategy that merges the high-dimensional embeddings from the same pool $\mathcal{P}$ into a joint latent representation. We concatenate the embeddings from the selected models along the feature dimension to form a unified feature vector:
\begin{equation}
h_{\text{concat}} = [h_{f}^\top, h_{1}^\top, h_{2}^\top, \dots, h_{K}^\top]^\top \in \mathbb{R}^{d (K+1)},
\end{equation}
where $d=32$ is the embedding dimension and $K$ is the number of selected subband models. The resulting vector is processed by an Multi-Layer Perceptron (MLP) with two hidden layers (256 and 128 units) to map these joint features to a logit $z_{\text{concat}}$. 
This architecture parameterized cross-expert correlations, enabling the MLP to learn non-linear dependencies and assign adaptive importance to global and localized spectral features. 

\subsubsection{Cross-Expert Interaction}
While the concatenation strategy relies on input-agnostic MLP mapping, Cross-Expert Interaction introduces a dynamic relational mechanism to capture signal-dependent forgery traces. Instead of static stacking, we employ Multi-Head Self-Attention (MHSA) \cite{vaswani2017attention} to allow fullband and subband experts to mutually refine their representations through cross-expert communication. We represent the pooled embeddings as a sequence $H = [h_f, h_1, h_2, \dots, h_{K}]^\top \in \mathbb{R}^{(K+1) \times d}$. The MHSA block refines these features by calculating pairwise affinities:
\begin{equation}
\text{Attention}(Q, K, V) = \text{softmax}\left(\frac{QK^\top}{\sqrt{d_k}}\right)V,
\end{equation}
where $Q = HW_Q$, $K = HW_K$, and $V = HW_V$ are learnable linear projections of $H$. 
$d_k$ is the dimensionality of each attention head, and the factor $1/\sqrt{d_k}$ prevents large dot-product magnitudes~\cite{DBLP:journals/corr/VaswaniSPUJGKP17}.
Attention-based mechanisms provide an effective means of aggregating information from diverse perspectives \cite{chen2025towards, chen2025localizing}. 
Unlike the static receptive field of an MLP, this mechanism captures non-local spectral dependencies by dynamically recalibrating the expert contributions based on its relevance to discriminative forgery cues. The refined sequence is then aggregated via global average pooling and passed through a shallow MLP head to produce the final logit $z_{\text{inter}}$. 
This mechanism enables dynamic prioritization of salient spectral fingerprints while suppressing noise from irrelevant subbands. 

\subsubsection{Cross-Expert Distillation}
While ensemble models and attention mechanisms improve performance, they significantly increase computational costs during inference. 
Cross-Expert Distillation addresses this by transferring specialized knowledge from subband experts into a single fullband student model. 
Unlike traditional distillation-based model compression \cite{hinton2015distilling, chen2024mtdvocalist, liao22_spsc}, our goal is to enhance the student model's capacity, teaching it to recognize specific frequency-localized patterns without increasing the final model size. 

\textbf{1) Knowledge Distillation Objectives. }
We employ two distillation objectives targeting both logit-level and feature-level knowledge. 
For a given input, the student produces logit $z^{(s)}$ and embedding $h^{(s)}$, while the teacher provides $z^{(t)}$ and $h^{(t)}$.

\begin{itemize}
    \item \textit{Logit-Level Knowledge. } We utilize Kullback-Leibler (KL) divergence~\cite{kullback1951information} to minimize the discrepancy between softened probability distributions:
    \begin{equation}
        \mathcal{L}_{\text{logit}} = \tau^2 D_\text{KL}(\sigma(z^s/\tau) || \sigma(z^t/\tau)),
    \end{equation} where $\sigma$ is the binary softmax function and $\tau$ is the temperature. This encourages the student to align its output distribution with the teacher’s softened predictions.

    \item \textit{Feature-Level Knowledge. } To internalize spectral sensitivity, we minimize the Mean Squared Error (MSE) between latent embeddings:
    \begin{equation}
        \mathcal{L}_{\text{feat}} = | h^s - h^t |^2_2.
    \end{equation} This ensures the student's latent space captures the specialized forgery fingerprints identified by the teacher.
\end{itemize}
The overall objective is defined as
$
\mathcal{L}_{\text{total}} 
= \mathcal{L}_{\text{ce}} 
+ \alpha \mathcal{L}_{\text{logit}} 
+ \beta \mathcal{L}_{\text{feat}},
$
where $\mathcal{L}_{\text{ce}}$ is the standard cross-entropy loss, and $\alpha$ and $\beta$ balance decision-level alignment and feature-level supervision. 

\textbf{2) Teacher Configurations. }
We propose a multi-teacher distillation framework to transfer knowledge from diverse subband experts to a fullband student model. This framework allows the student to integrate specialized subband knowledge into a unify representation. 
The aggregated teacher embedding $h^{(t)}$ and logit $z^{(t)}$ are defined as:
\begin{equation}
h^{(t)} = \sum_{m=1}^{M} w_m h^{(t_m)}, \quad z^{(t)} = \sum_{m=1}^{M} w_m z^{(t_m)},
\end{equation}
where $t_m$ is the $m$-th subband expert for a specific frequency region, and $w_m$ are weights satisfying $\sum_{m=1}^{M} w_m = 1$. The fullband student is optimized to minimize the discrepancy between its global outputs and these aggregated subband targets. 
Based on above framework, we implement two distinct teacher configurations for the fullband student model:
\begin{itemize}
    \item \textit{Single-Teacher Distillation ($M=1$). }
    By learning from a single subband expert, the fullband student intensifies its focus on internalizing specific, localized artifacts identified by the specialized teacher within a targeted frequency range. 
    \item  \textit{Dual-Teacher Distillation ($M=2$). } Guided by two non-overlapping subband experts, the student broadens its learning scope to reconcile different frequency cues. This objective encourages the fullband model to integrate global consistency with the diverse local details provided by both teachers. 
\end{itemize}
By adopting this Cross-Expert Distillation approach, the fullband student model effectively digests ensemble knowledge, achieving a compact and efficient architecture without sacrificing the specialized insights from the subband experts. 

\section{Experimental Setup}\label{sec:exp_setup}
To evaluate the performance of Sing-HiResNet framework, this section details the dataset, evaluation protocols, pre-processing procedures, model setup, and distillation configurations.  

\textbf{Dataset and Evaluation.}
We evaluate our model on the WildSVDD dataset \cite{zhang2024svdd}, which contains authentic and AI-synthesized singing from unconstrained online sources. The dataset comprises recordings from a total of 97 singers with 3,223 songs. 
Due to copyright, we re-collected the data following official protocols, resulting in minor distribution shifts. The training set comprises 27,879 utterances (15,364 deepfake / 12,515 bonafide), with 20\% reserved for validation. We evaluate performance on Test A, featuring unseen singers in the training language (2,774 / 2,849), and Test B, consisting of unseen Persian singers from an out-of-distribution language (91 / 236).  
Given that instrumental accompaniment provides negligible discriminative cues \cite{chen2025doesinstrumentalmusichelp}, we focus exclusively on vocal-only SingFake detection in this paper. 
All model performance was measured using the pooled equal error rate (EER).

\textbf{Pre-processing and Model Setup. }
We adopted the data preparation procedure provided by UNIBS~\cite{10888452}. All audio clips are standardized to a 4-second duration through cropping or repeat padding at a 44.1 kHz sampling rate. These audio clips are then converted into log-power spectrograms to serve as the model input. The backbone is a ResNet-18, pre-trained on ImageNet \cite{imagenet} and modified to accommodate single-channel spectrograms. To adapt the architecture for subband feature extraction, we replace the original fully connected layer with a compact projection head that produces a 32-dimensional embedding for each segment. 
For optimization, training is conducted using sigmoid focal loss \cite{lin2017focal}, with the AdamW \cite{loshchilov2017decoupled} optimizer and a CosineAnnealingLR \cite{loshchilov2016sgdr} scheduler. 

\textbf{Cross-Expert Distillation Setup. }
To facilitate logit-level distillation, we set the temperature $\tau = 3.0$ to soften the probability distributions. The overall training objective is controlled by balancing coefficients $\alpha = 0.5$ for logit-level and $\beta = 0.2$ for feature-level distillation, serving as an trade-off between supervised learning and knowledge transfer. For the dual-teacher configuration, we assign aggregation weights of $w_1 = 0.6$ and $w_2 = 0.4$ to the low-frequency and mid/high-frequency experts, respectively. This distribution prioritizes the low-frequency band, which our empirical analysis identified as the primary contributor to overall performance. 

\section{Experiment Results}

\begin{table}[t]
\centering
\fontsize{8}{10}\selectfont 
\setlength\tabcolsep{7.0pt} 
\renewcommand{\arraystretch}{1.3} 
\caption{EER (\%) of subband expert models across different subband partition configurations. Model names indicate approximate frequency ranges for readability, while the subband (kHz) column reports the exact band boundaries. \texttt{SB-Concat-$N$} denotes the feature-level concatenation model combining all subbands in the same configuration.}
\label{tab:submodel}

\begin{tabular}{lcccc}
\toprule
\textbf{Model} & \textbf{Subband (kHz)} & \textbf{Valid} & \textbf{Test A} & \textbf{Test B} \\
\midrule
\rowcolor{gray!15} \multicolumn{5}{l}{\textbf{Fullband $N=1$}} \\
\texttt{FB$_{[0.0, 22.0]}$} & 0.00--22.05 & 0.66 & 2.31 & 10.79 \\
\midrule
\midrule

\rowcolor{gray!15} \multicolumn{5}{l}{\textbf{Subband $N=2$}} \\
\rowcolor{blue!5} \texttt{SB$_{[0.0, 11.0]}$} & 0.00 -- 11.03 & 1.39 & 2.81 & 22.01 \\
\rowcolor{blue!5} \texttt{SB$_{[11.0, 22.0]}$} & 11.03 -- 22.05 & 1.71 & 5.66 & 15.32 \\
\texttt{SB-Concat-2} &  & 0.09 & 3.13 & 16.29 \\
\midrule
\midrule

\rowcolor{gray!15} \multicolumn{5}{l}{\textbf{Subband $N=4$}} \\
\texttt{SB$_{[0.0, 5.5]}$}  & 0.00 -- 5.51 & 2.69 & 6.28 & 27.51 \\
\texttt{SB$_{[5.5, 11.0]}$} & 5.51 -- 11.03 & 3.04 & 5.05 & 20.82 \\
\rowcolor{blue!5} \texttt{SB$_{[11.0, 16.5]}$} & 11.03 -- 16.54 & 1.45 & 4.46 & 17.48 \\
\texttt{SB$_{[16.5, 22.0]}$} & 16.54 -- 22.05 & 3.47 & 16.08 & 29.45 \\
\texttt{SB-Concat-4}          &  & 0.22 & 3.54 & 17.48 \\
\midrule
\midrule
\rowcolor{gray!15} \multicolumn{5}{l}{\textbf{Subband $N=8$}} \\
\texttt{SB$_{[0.0, 2.8]}$} & 0.00--2.76 & 2.86 & 8.07 & 28.48 \\
\texttt{SB$_{[2.8, 5.5]}$} & 2.76--5.51 & 4.40 & 11.06 & 30.85 \\
\texttt{SB$_{[5.5, 8.3]}$} & 5.51--8.27 & 6.45 & 11.79 & 30.85 \\
\texttt{SB$_{[8.3, 11.0]}$} & 8.27--11.03 & 4.18 & 6.85 & 35.17 \\
\texttt{SB$_{[11.0, 13.8]}$} & 11.03--13.78 & 3.60 & 6.99 & 16.50 \\
\texttt{SB$_{[13.8, 16.5]}$} & 13.78--16.54 & 3.95 & 9.09 & 26.32 \\
\texttt{SB$_{[16.5, 19.3]}$} & 16.54--19.29 & 5.67 & 20.22 & 31.82 \\
\texttt{SB$_{[19.3, 22.0]}$} & 19.29--22.05 & 7.62 & 24.36 & 27.51 \\
\texttt{SB-Concat-8} &  & 0.18 & 3.17 & 22.98 \\
\bottomrule
\end{tabular}
\end{table} 

\subsection{Preliminary Study of Subband Modeling}
\label{sec:subband_modeling_res}
Table~\ref{tab:submodel} evaluates the efficacy of subband expert models across four partition configurations ($N=1, 2, 4, 8$). To further investigate the potential of these experts, we also evaluate feature-level concatenation variants (\texttt{SB-Concat-$N$}), which merge the embeddings of all subband experts within a partition without the fullband model. These results allow us to analyze how localized spectral information contributes to detection performance both individually and as an ensemble. Based on this evaluation, we derive several key insights regarding the spectral distribution of spoofing artifacts in singing voices. 

\textbf{The Necessity of the Fullband Expert Model. } A critical question is whether an ensemble of non-overlapping subband experts can effectively replace a single model trained on the fullband spectrum. While Table~\ref{tab:submodel} shows that \texttt{SB-Concat-$N$} models often achieve superior results on the validation set, they exhibit significant performance degradation in both in-domain and out-of-domain evaluations. This suggests that while feature-level concatenation captures localized cues, the subband slicing strategy might destroy some of the critical cues around the subband cutting boundaries. Moreover, without the global guidance of a fullband model, the ensemble lacks the holistic perspective necessary for robust generalization, leading to the ignorance of the actual distribution of discriminative artifacts. Consequently, the fullband expert remains critical as it provides the global context required to balance and calibrate the varying importance of individual subband insights.

\textbf{The Impact of Subband Modeling. }
The performance of expert models varies significantly depending on the spectral granularity, revealing a critical trade-off between artifact isolation and feature sufficiency. 
On the one hand, results show that spoofing artifacts are distributed unevenly across the spectrum. When in-domain, the lower subband $\texttt{SB}_{[0.0, 11.0]}$ (1.39\% EER) and the mid-high subband $\texttt{SB}_{[11.0, 16.5]}$ (1.45\% EER) are the most effective experts. However, their generalization performance differs: $\texttt{SB}_{[0.0, 11.0]}$ performs best on Test A (2.81\%), while the higher-frequency experts $\texttt{SB}_{[11.0, 22.0]}$ (15.32\%) and $\texttt{SB}_{[11.0, 16.5]}$ (17.48\%) are more robust on Test B. This suggests that while low frequencies provide a strong baseline, the 11.0–22.0 kHz range contains vital, localized cues that are key to handling domain shifts, especially when broader frequency windows might otherwise hide these subtle details.
On the other hand, excessively fine-grained splitting can be counterproductive. When the spectrum is divided into too many narrow bands (e.g., $N=8$), the information within each subband becomes too limited to support reliable deepfake detection. For instance, in the $N=8$ configuration, $\texttt{SB}_{[19.3, 22.0]}$ yields the worst performance on Test A (24.36\% EER), while $\texttt{SB}_{[8.3, 11.0]}$ struggles most on Test B with a high EER of 35.17\%. This indicates that while isolation is helpful, a subband must still retain enough spectral context to capture meaningful artifacts; otherwise, the expert model fails to learn robust discriminative features. 

Our preliminary study indicates that while fullband models remain highly effective at capturing global contextual information for SingFake detection, high-resolution audio often contains localized artifact cues within specific subbands. Based on these observations, there is significant potential to enhance model performance by amplifying key subband artifact cues while preserving a global perspective. Accordingly, we propose Sing-HiResNet in Section~\ref{sec:Sing-HiResNet}, a framework that leverages joint fullband-subband modeling to achieve this synergy. 

\begin{table*}[t]
\centering
\fontsize{8}{10}\selectfont 
\setlength\tabcolsep{8pt} 
\renewcommand{\arraystretch}{1.3} 
\caption{EER (\%) of the proposed integration framework across different component configurations.}
\label{tab:subband_integrate_modified}

\begin{tabular}{lccccccccccccc} 
\toprule
& \multicolumn{4}{c}{Selected Model Pool $\mathcal{P}$} & \multicolumn{2}{c}{Aggregation} & \multicolumn{2}{c}{Concatenation} & \multicolumn{2}{c}{Interaction} & \multicolumn{2}{c}{Distillation\textsuperscript{*}} \\

 \cmidrule(lr){2-5} \cmidrule(lr){6-7} \cmidrule(lr){8-9} \cmidrule(lr){10-11} \cmidrule(lr){12-13}
&\texttt{FB} & \texttt{SB$_\texttt{L}$} & \texttt{SB$_\texttt{M}$} & \texttt{SB$_\texttt{H}$} & Test A & Test B & Test A & Test B & Test A & Test B & Test A & Test B \\
\midrule

Pool $\mathcal{P}_a$ & \checkmark & \checkmark & & & \textbf{1.73} & 10.79 & 2.95 & 15.53 & \textbf{2.99} & 15.32 & 1.83 & 11.00 \\
Pool $\mathcal{P}_b$ & \checkmark & & & \checkmark & 2.56 & 11.21 & 2.95 & 20.82 & 3.24 & 16.29 & 3.13 & 12.19 \\
Pool $\mathcal{P}_c$ & \checkmark & & \checkmark & & 2.28 & 10.79 & \textbf{2.85} & \textbf{16.50} & 3.02 & 15.32 & 2.17 & 10.79 \\
\midrule

Pool $\mathcal{P}_d$ & \checkmark & \checkmark & & \checkmark & 1.87 & 9.82  & 3.57 & 19.64 & 3.02 & 16.50 & 1.87 & 12.19 \\
Pool $\mathcal{P}_e$ & \checkmark & \checkmark & \checkmark & & \textbf{1.73} & \textbf{8.84} & 3.43 & 24.16 & 3.75 & \textbf{7.45} & \textbf{1.65} & \textbf{9.06} \\
\bottomrule
\addlinespace[2pt]
\multicolumn{13}{l}{\scriptsize \textit{* During knowledge distillation, the fullband model ($FB$) serves as the student model, while the subband models ($SB_{\sim}$) act as the teacher models.}}
\end{tabular}
\end{table*}

\begin{figure*}[t]
    \centering
    \includegraphics[width=17cm]{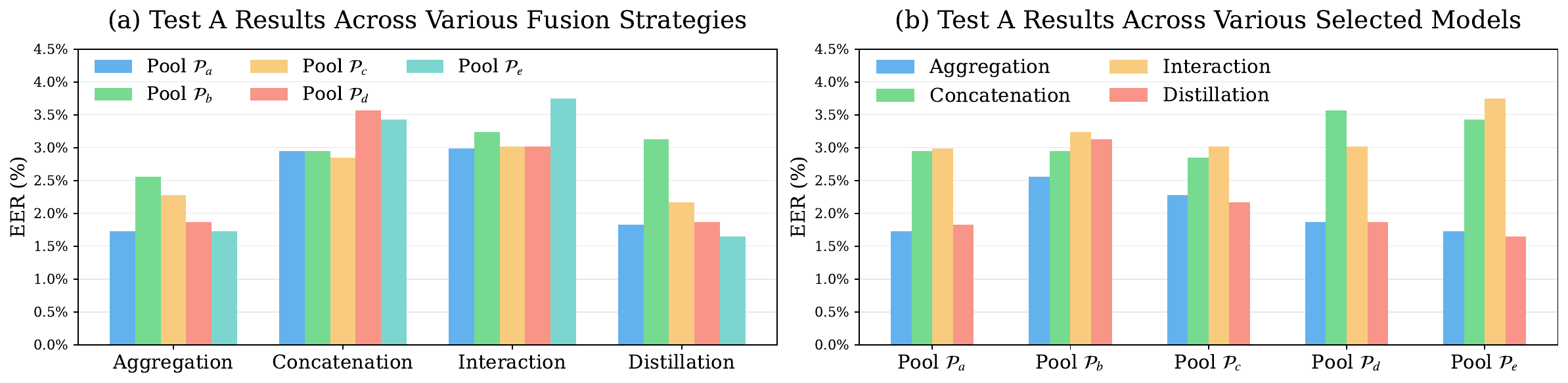}
    \includegraphics[width=17cm]{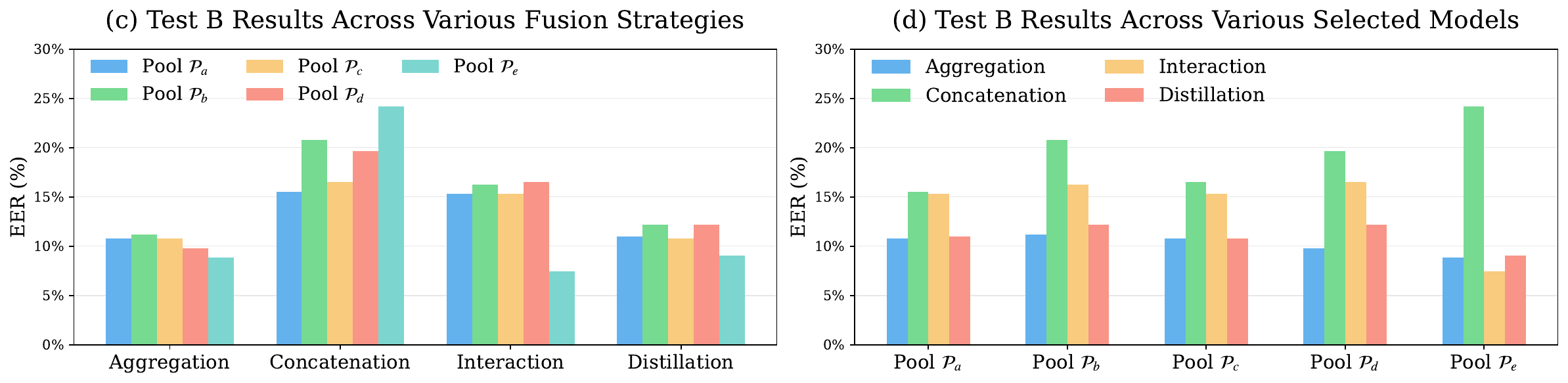}
    \caption{EER (\%) results across two categorization schemes. The left columns present a method-centric view to highlight frequency impact, while the right columns provide a condition-centric comparison of integration strategies across Test A and Test B.}
    \label{fig:integrate}
\end{figure*}

\begin{figure*}[t]
\centering
\includegraphics[width=17cm]{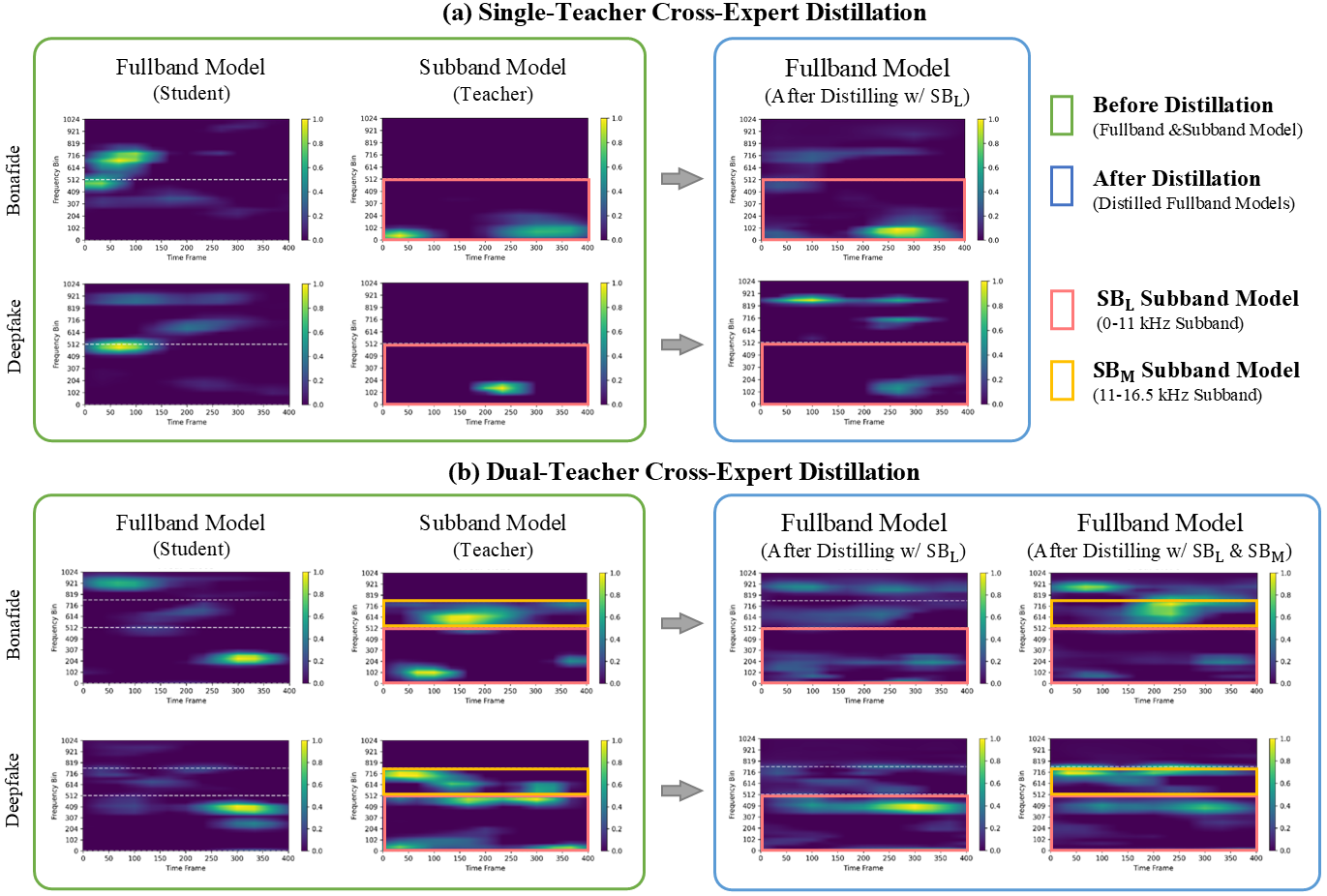}
\caption{Grad-CAM visualizations of expert and distilled models for bonafide and deepfake samples, featuring (a) single-teacher (Low) and (b) dual-teacher (Low/Mid-High) distillation. White dashed lines marking corresponding subband boundaries.
}
\label{fig:visualize}
\end{figure*}

\subsection{Comparative Analysis of Fusion Strategies}
\label{sec:fusion_strategies}

To fully capture both global patterns and localized artifacts, we evaluate how to best combine different spectral cues. While the fullband expert ($\texttt{FB}$) focuses on the overall spectrum, the subband experts provide more detail in specific subbands. Table~\ref{tab:subband_integrate_modified} compares four fusion strategies across model pools $\mathcal{P}_a$--$\mathcal{P}_e$. These strategies integrate a fullband model with three specialized subband models: $\texttt{SB}_\texttt{L}$ (Low: 0.0--11.03 kHz), $\texttt{SB}_\texttt{M}$ (Mid-High: 11.03--16.5 kHz), and $\texttt{SB}_\texttt{H}$ (High: 11.0--22.05 kHz). 
To evaluate the optimal synergy between fullband and subband modeling, we analyze various fusion strategies and subband configurations across the experimental model pools.  
Figures~\ref{fig:integrate}(a) and (c) compares the performance of integration strategies across model pools $\mathcal{P}_a$ to $\mathcal{P}_e$. Overall, Decision-Level Aggregation and Cross-Expert Distillation consistently outperform both Feature-Level Concatenation and Cross-Expert Interaction mechanisms across most model pools. 
Although Interaction (via MHSA) excels on the out-of-distribution Test B ($\mathcal{P}_e$), Aggregation provides a more reliable overall baseline by merging logits directly. 
The choice of subband components also significantly impacts the resulting EERs as illustrated in Figures~\ref{fig:integrate}(b) and (d). 
Within the most efficient joint fullband-subband fusion strategy (aggregation and distillation), the selected model pool $\mathcal{P}_e$ consistently achieves the lowest EER. 
As shown in Table~\ref{tab:subband_integrate_modified}, the model pool $\mathcal{P}_e$ includes only the subband experts $\texttt{SB}_\texttt{L}$ and $\texttt{SB}_\texttt{M}$, excluding $\texttt{SB}_\texttt{H}$. 
This suggests that incorporating $\texttt{SB}_\texttt{H}$ (11.0--22.05 kHz) yields no incremental performance gain during the fusion process. 

To understand the reasons behind above observations, we analyze subband feature integration through the lenses of representation integrity and spectral relevance. Architecturally, while Feature-Level Concatenation and Cross-Expert Interaction leverage direct supervision, Decision-Level Aggregation and Cross-Expert Distillation prioritize preserving independent expert judgments. However, as noted in Section~\ref{sec:subband_modeling_res}, information within a single subband is often too sparse for reliable SingFake detection. This scarcity leaves experts vulnerable to non-discriminative noise, especially in ultrahigh frequency regions. While hard labels can train experts to ignore such noise during score prediction, their underlying feature representations often remain ``tainted'' by this confusing information, ultimately undermining the model's overall diagnostic integrity. 

Beyond structural constraints, the physical distribution of artifacts further complicates integration, as discriminative cues concentrate primarily below 16.5 kHz. Consequently, the $\texttt{SB}_\texttt{H}$ (11.0–22.05 kHz) subband is notably less informative, echoing psychoacoustic findings by Ashihara (2007) \cite{ashihara2007hearing} regarding the sharp decline in human hearing sensitivity beyond 16 kHz. Because generative models prioritize reconstruction fidelity within audible ranges, higher frequencies are frequently under-modeled due to the resolution bottlenecks of Mel-spectrograms and the lack of inductive bias in transposed convolutions, which often trigger aliasing artifacts \cite{yang2025enhancing, lee2023bigvgan} that often lack consistent forensic signatures and are easily masked by stochastic noise.

In summary, by strategically prioritizing $\texttt{SB}_\texttt{L}$ and $\texttt{SB}_\texttt{M}$ over noisier $\texttt{SB}_\texttt{H}$, the fusion process can bypass ultrahigh frequency interference and instead leverage the specific spectral inconsistencies where generative models most noticeably struggle. 

\begin{table*}[t]
\centering
\fontsize{8}{11}\selectfont 
\setlength\tabcolsep{11.5pt}
\caption{EER (\%) of cross-integration strategies. \texttt{SB$_\texttt{L}$} and \texttt{SB$_\texttt{M}$} denote expert models for 0--11 kHz and 11--16.5 kHz bands, respectively. 
}
\label{tab:cross-integrate-new}
\begin{tabular}{llllcccccc}
\toprule
& \multirow{2}{*}{Notation} &  \multirow{2}{*}{Description} & \multirow{2}{*}{Methods}  & \multicolumn{4}{c}{Selected Model Pool $\mathcal{P}$} & \multicolumn{2}{c}{EER (\%)} \\

\cmidrule(lr){5-8} \cmidrule(lr){9-10}
 &  &  &  &  \texttt{FB$_\texttt{D-LM}$} & \texttt{FB} & \texttt{SB$_\texttt{L}$} & \texttt{SB$_\texttt{M}$} & Test A & Test B \\

\midrule
(a) & \texttt{FB$_\texttt{D-LM}$} & Pool $\mathcal{P}_e$ in Table~\ref{tab:subband_integrate_modified} & Distillation \textsuperscript{*}  & & \checkmark & \checkmark & \checkmark & 1.65 & 9.06 \\
(b) & \texttt{FB$_\texttt{A-LM}$} & Pool $\mathcal{P}_e$ in Table~\ref{tab:subband_integrate_modified} & Aggregation  & & \checkmark & \checkmark & \checkmark & 1.73 & 8.84 \\
\midrule
\rowcolor{gray!10}
(c) & \texttt{FB$_\texttt{SA-D-LM}$}  & Subband & Aggregation & \checkmark & & \checkmark & \checkmark & 1.73 & 12.19 \\
\rowcolor{gray!10}
(d) & \texttt{FB$_\texttt{FSA-D-LM}$} & Fullband-Subband & Aggregation & \checkmark & \checkmark & \checkmark & \checkmark & \textbf{1.58} & \textbf{8.77} \\
\bottomrule
\addlinespace[2pt]
\multicolumn{8}{l}{\textit{\scriptsize * \texttt{FB} is the fullband student model, \texttt{SB$_{\sim}$} are the subband expert teacher models.}}
\end{tabular}
\end{table*}

\subsection{Evidence of Subband-Aware Knowledge Transfer}
\label{exp:evidence}

To verify whether distillation effectively transfers frequency-localized expertise, we perform a visual attribution analysis using Grad-CAM \cite{selvaraju2020grad}. 
We apply it to the final convolutional layers of the ResNet18 backbone. We compute the input log-power spectrogram via a standard forward pass and backpropagate gradients from the target logit to the feature maps. This process generates activation-weighted importance maps, which are normalized and overlaid onto the spectrogram. These heatmaps highlight the specific time–frequency regions the model prioritizes during classification. 
Our analysis focuses on four samples from Test A, which in Figure~\ref{fig:visualize}. 
These samples were misclassified by the baseline fullband model but correctly identified after distillation. By visualizing these four cases, we demonstrate how the student model successfully resolves these misclassified cases by shifting its attention to the relevant spectral subbands. 

We compare the attention heatmaps before and after distillation to examine how the student adapts its focus according to different teacher configurations. 
Figure~\ref{fig:visualize}(a) illustrates the effects of Single-Teacher Cross-Expert Distillation. We observe that the original fullband student model tends to focus excessively on high-frequency regions, regardless of whether the input is bonafide or deepfake. In contrast, the subband teacher model ($\texttt{SB}_\texttt{L}$) primarily attends to the low-frequency regions for both classes. Following distillation from the teacher model $\texttt{SB}_\texttt{L}$, the student model successfully redirects its focus toward the low-frequency regions. 
Similarly, Figure~\ref{fig:visualize}(b) illustrates the effects of Dual-Teacher Cross-Expert Distillation. We observe that the original fullband student model's attention is biased toward the lower-right regions of the log-spectrogram for both bonafide and deepfake samples. Notably, it largely disregards the critical spectral regions prioritized by the subband teachers ($\texttt{SB}_\texttt{L}$ and $\texttt{SB}_\texttt{M}$). 
Following distillation from the subband teacher models ($\texttt{SB}_\texttt{L}$ and $\texttt{SB}_\texttt{M}$), the student model successfully redirects its focus to encompass the informative regions from both subband teacher models, effectively bridging the knowledge gap between the fullband and subband specialized expert models, leading to a more comprehensive feature representation.

In summary, these observations provide qualitative evidence that the proposed cross-expert distillation successfully facilitates the transfer of frequency-localized expertise. By aligning its attention with the specialized teachers, the student model learns to capture discriminative spectral cues that were previously overlooked by the fullband model, ultimately leading to more robust classification on challenging samples. 

\subsection{Synergizing Knowledge Distillation and Score Fusion}
Since Section~\ref{sec:fusion_strategies} shows that both cross-expert distillation and decision-level aggregation improve the EER on Test A and Test B, we select these two strategies for further integration. 
Table~\ref{tab:cross-integrate-new} examines if combining distillation with aggregation can further enhance performance. We treat the distilled student (\texttt{FB$_\texttt{D-LM}$}) as an additional expert model and add it to the model pool to test this combination. 
Rows (a) and (b) represent the original distillation and aggregation strategies using the model pool $\mathcal{P}_e$ from Table~\ref{tab:subband_integrate_modified}. Rows (c) and (d) further apply decision-level aggregation by combining the distilled student (\texttt{FB$_\texttt{D-LM}$}) with additional subband or fullband-subband expert models. 

Experimental results in Table~\ref{tab:cross-integrate-new} demonstrate that treating the distilled student as an additional expert consistently improves performance. Notably, the fullband–subband aggregation (row d) achieves the lowest EER of 1.58\% on Test A and also yields the best performance on Test B with an EER of 8.77\%. These results indicate that the distillation-enhanced integration strategy strengthens cross-expert complementarity and improves robustness not only to unseen singers within the same language but also under cross-lingual conditions. 

\begin{table}[t]
\centering
\fontsize{7.5}{10}\selectfont
\setlength\tabcolsep{3pt} 
\renewcommand{\arraystretch}{1.1}

\caption{Performance comparison on WildSVDD dataset, where $f_s$ denotes sampling rate (kHz). Results are presented in Equal Error Rate (EER, \%) with 95\% Confidence Intervals (CI). }
\label{tab:comparison}

\begin{tabular}{lllll}
\toprule
Models & Backbone & \boldmath{$f_s$} & Test A (CI) & Test B (CI) \\
\midrule
Baseline1 & Wav2vec         & 16   & 6.09 & 24.09 \\
Baseline2 & Raw             & 16   & 8.84 & 26.11 \\
\midrule
PDL       & Ensemble        & 16   & 5.80 & 22.01 \\
NTU       & SingGraph       & 16   & 4.31 & 31.82 \\
IMS-SCU   & WavLM           & 16   & 3.54 & 15.32 \\
IMS-SCU   & Ensemble        & 16   & 2.70 & 12.95 \\
UNIBS     & ResNet18        & 44.1 & 2.38 & \phantom{0}9.81 \\
\midrule
\rowcolor{gray!10}
Our UNIBS & ResNet18           & 44.1 & 2.31 {\tiny (1.94--2.81)} & 10.79 {\tiny (6.39--15.60)} \\ 
\rowcolor{gray!10}
Sing-HiResNet & \texttt{FB$_\texttt{D-LM}$}  & 44.1 & \underline{1.65} {\tiny (1.35--2.06)} & \phantom{0} 9.06 {\tiny (6.11--14.09)} \\ 
\rowcolor{gray!10}
Sing-HiResNet & \texttt{FB$_\texttt{A-LM}$}  & 44.1 & 1.73 {\tiny (1.35-2.06)} & \phantom{0} 8.84 {\tiny (5.54--15.01)} \\ 
\rowcolor{gray!10}
Sing-HiResNet & \texttt{FB$_\texttt{I-LM}$}  & 44.1 & 3.75 {\tiny (3.25--4.29)} & \phantom{0} \textbf{7.45} {\tiny (4.33--10.59)} * \\ 
\rowcolor{gray!10}
Sing-HiResNet & \texttt{FB$_\texttt{FSA-D-LM}$} & 44.1 & \textbf{1.58} {\tiny (1.26--1.94)} * & \phantom{0} \underline{8.77} {\tiny (7.15--12.54)} \\
\bottomrule
\addlinespace[2pt]
\multicolumn{5}{p{0.95\columnwidth}}{\textit{\scriptsize * The best models achieve relative EER reductions of 31.6\% and 30.9\% on Test A and Test B, respectively, compared to the re-implemented UNIBS. }}
\end{tabular}
\end{table}

\subsection{Benchmarking Against State-of-the-Art Systems}
Table~\ref{tab:comparison} compares our proposed Sing-HiResNet models with top-performing systems from the SVDD Challenge 2024. Baselines 1 and 2 represent the official model baselines, while the remaining results are referenced from the official challenge leaderboards \cite{svdd2024leaderboard}.
The audio sampling rate of most competitive approaches is 16~kHz, which inevitably filters out critical high-frequency cues. In contrast, UNIBS is the only existing baseline that utilizes a 44.1~kHz sampling rate. To ensure a rigorous and fair comparison, we independently re-implemented the UNIBS architecture as our primary baseline under identical experimental conditions. 
Consequently, based on the superior performances results in Table~\ref{tab:subband_integrate_modified}, 
we specifically selected several representative model based on pool $\mathcal{P}_e$ featuring three effective fusion strategies: distillation, aggregation, and interaction. 

As shown in Table~\ref{tab:comparison}, large-scale SSL-based models (e.g., WavLM, Wav2Vec) exhibit significant performance degradation on Test B compared to Test A. This suggests that while SSL models excel at capturing general phonetic features, they may lack the high-frequency sensitivity required for ``in-the-wild'' singing voice forensics. 
In contrast, our Sing-HiResNet maintains more stable EERs across both test sets. This confirms that explicitly preserving high-resolution spectral content is more effective for generalizing to unconstrained environments than relying on massive pre-training at audio data with low-resolution. Overall, our proposed variants, including $\texttt{FB}_\texttt{D-LM}$, $\texttt{FB}_\texttt{A-LM}$, and $\texttt{FB}_\texttt{FSA-D-LM}$, consistently outperform the re-implemented UNIBS on both Test A and Test B. Notably, $\texttt{FB}_\texttt{FSA-D-LM}$ achieves the best performance on Test A (1.58\% EER), while $\texttt{FB}_\texttt{I-LM}$ leads on Test B (7.45\% EER). These results indicate that while specific strategies excel in different evaluation scenarios, achieving a single architecture that dominates both Test A and Test B remains an ongoing challenge. 

\section{Conclusion}
This study provides the first systematic analysis of joint fullband-subband modeling for high-resolution SingFake detection by leveraging audio input at a 44.1 kHz sampling rate. 
We argue that high-resolution audio at 44.1 kHz preserves extended harmonics and breath textures essential for forgery detection, whereas audio downsampled to 16 kHz discards vital cues.
Motivated by these insights, we introduced Sing-HiResNet, a framework that concurrently captures global spectral patterns and frequency-specific artifacts. Among four fusion strategies, Decision-Level Aggregation and Cross-Expert Distillation most effectively leveraged the synergy between scales. Grad-CAM visualizations confirm that our distillation strategy successfully transfers localized expertise to a holistic fullband student model. 
Sing-HiResNet achieves state-of-the-art results on the WildSVDD dataset, reaching an EER of 1.58\% on Test A and 7.45\% on Test B. These results represent relative EER reductions of 31.6\% and 30.9\% over the baseline. These findings establish joint fullband-subband modeling as a critical requirement for robust, in-the-wild singing voice forensics. 

\section{Acknowledgements}
This work was supported by the Ministry of Education (MOE) of Taiwan under the project "Taiwan Centers of Excellence in Artificial Intelligence," through the NTU Artificial Intelligence Center of Research Excellence. We acknowledge the National Center for High-performance Computing (NCHC) for providing essential computational resources. Additionally, the authors are grateful to Prof. Zhiyao Duan at the University of Rochester for his insightful feedback on the experimental design. 

\section{Generative AI Use Disclosure}
We employed Gemini for grammatical paraphrasing and language polishing to improve the manuscript's clarity. 
The AI tool was utilized solely for technical editing purposes and did not contribute to the conceptualization, data analysis, or production of any significant scholarly content in this work. 

\bibliographystyle{IEEEtran}
\bibliography{refs}

\end{document}